# Wide Stiffness Range Cavity Optomechanical Sensors for Atomic Force Microscopy

Yuxiang Liu,[1,2] Houxun Miao,[1,3] Vladimir Aksyuk,[1,*] and Kartik Srinivasan[1,*]

[1]*Center for Nanoscale Science and Technology, National Institute of Standards and Technology, Gaithersburg, Maryland 20899, USA*
[2]*Institute for Research in Electronics and Applied Physics, University of Maryland, College Park, Maryland 20742*
[3]*Maryland Nanocenter, University of Maryland, College Park, Maryland 20742, USA*
[*]*kartik.srinivasan@nist.gov; vladimir.aksyuk@nist.gov*

**Abstract:** We report on progress in developing compact sensors for atomic force microscopy (AFM), in which the mechanical transducer is integrated with near-field optical readout on a single chip. The motion of a nanoscale, doubly-clamped cantilever was transduced by an adjacent high quality factor silicon microdisk cavity. In particular, we show that displacement sensitivity on the order of 1 fm/(Hz)$^{1/2}$ can be achieved while the cantilever stiffness is varied over four orders of magnitude ($\approx$ 0.01 N/m to $\approx$ 290 N/m). The ability to transduce both very soft and very stiff cantilevers extends the domain of applicability of this technique, potentially ranging from interrogation of microbiological samples (soft cantilevers) to imaging with high resolution (stiff cantilevers). Along with mechanical frequencies (>250 kHz) that are much higher than those used in conventional AFM probes of similar stiffness, these results suggest that our cavity optomechanical sensors may have application in a wide variety of high-bandwidth AFM measurements.

**1. Introduction**

Scanning force microscopy, especially atomic force microscopy (AFM), has been an essential tool for micro-/nanoscale studies in physics, chemistry, and biology, thanks to its ability to characterize the topography of a surface down to the level of individual atoms [1, 2] and to locally measure very small surface forces. Although conventional AFM is well-developed, progress in MEMS and nanophotonics may provide performance improvements. For example, monolithic integration of the mechanical transducer (the AFM cantilever) and the optical readout scheme within a single silicon chip can make the system compact, stable, and compatible with low-cost, batch fabrication. Developments in cavity optomechanics [3-5] suggest that optical readout can be done with high sensitivity through near-field coupling to optical microresonators. Within such a scheme, the cantilever size can be shrunk to nanoscale dimensions, a size regime that is difficult to effectively transduce through free-space optical techniques due to strong diffraction effects, which occur when the cantilever width is smaller than the detection beam waist, and which compete with the reflection of the detection laser at the cantilever tip and hence limit the AFM sensitivity [1, 6]. Since nanoscale cantilevers can combine high (MHz) resonance frequencies with moderate stiffness ($k \approx 1$ N/m), their effective utilization may help increase the image acquisition speed and/or time-resolution of AFM. We recently demonstrated a silicon cavity optomechanical system designed with AFM applications in mind [7], in which a semicircular, doubly-clamped nanoscale cantilever is held within the near-field of a high quality factor ($Q$) microdisk cavity (Fig. 1a). Thermally-driven motion of the 2.35 MHz fundamental mode of the cantilever was transduced with a displacement sensitivity of $\approx (4.4 \pm 0.3) \times 10^{-16}$ m/(Hz)$^{1/2}$.

The central element of an AFM is the cantilever spring that transduces the sample-tip interaction force [2]. One of the most important parameters of the cantilever is its normal spring constant $k$. Typically, there is a trade-off between the force sensitivity and the system stability, both of which depend on the value of $k$. On the one hand, to ensure high force sensitivity, $k$ should be small, so that the ratio of the mechanical displacement signal to displacement readout noise is maximized, ideally with the cantilever thermal noise dominating the readout noise over the full bandwidth of the measurement. On the other hand, for static or small amplitude dynamic operation, the tip loses stability and jumps into contact with the sample when the positive gradient of the interaction force is larger than $k$, so the system becomes less stable with a smaller $k$ and at a smaller separation distance. The gradient of interaction force varies strongly depending on the sample and distance, ranging from $\approx 0.1$ N/m for biological samples to $\approx 100$ N/m for solids [8]. Therefore, the cantilever stiffness has to be carefully chosen for different samples and applications. The general stiffness range of conventional AFM cantilevers is from $\approx 0.01$ N/m to $\approx 100$ N/m [1], although much softer and stiffer custom cantilevers have been reported in literature for special applications [2, 9].

In our previous work, the cantilever stiffness was limited to a range of $k \approx 0.1$ N/m to 4 N/m. It has also been suggested (Ref. [10]) that the curved cantilever geometry may limit the accessible range of the device size and mechanical properties, especially at the higher end of frequency and stiffness range. In this paper, we investigate the cantilever stiffness range that is compatible with our cavity optomechanical readout scheme. By changing the cantilever length and corresponding microdisk diameter (from 2.5 µm to 50 µm), as well as the cantilever width, we experimentally demonstrate a cantilever stiffness ranging from $\approx 0.01$ N/m to $\approx 290$ N/m, which covers the stiffness range of conventional AFM. Thermally-driven cantilever vibrations are transduced with a displacement sensitivity at the fm/(Hz)$^{1/2}$ (or better) level across the stiffness range. In addition, our small devices have much higher natural frequencies ($\approx 200$ kHz to $\approx 110$ MHz) than those used in conventional AFM (10 kHz to 1.1 MHz [8]), which may help reduce noise and improve image acquisition speed. These results indicate that our integrated sensing platform based on cavity optomechanics may be appropriate for a wide range of AFM applications, ranging from interrogating biological samples (on the soft side of the stiffness range) to imaging with a sub-atomic resolution (on the hard side of the stiffness range).

## 2. Working principle

The device consists of a semicircular cantilever curved around the edge of a waveguide-coupled microdisk optical resonator with a separation of $\approx 100$ nm, as shown in Fig. 1(a). The cantilever deformation and the resonant optical mode shown in Fig. 1(a) were calculated with a commercial finite element method (FEM) software (See Section 4.2 for more details). The cantilever is clamped at its two ends, which are far away from the disk to minimize their influence on it. When the cantilever moves with respect to the disk (due to thermal noise, for example), an optical wave circulating around the disk edge feels the resulting change in peripheral refractive index, producing an effective path length change for the microdisk's resonant optical modes which modulates their spectral position.  This is converted to an intensity modulation by coupling a light field from an input optical waveguide into the device, with the input field's wavelength tuned to the shoulder of one the microdisk cavity modes (see Fig. 1(c)) [3, 4]. The intensity modulated output field from the disk is out-coupled through the same waveguide and detected, and the electric signal from the detector is spectrally analyzed, with the cantilever's mechanical modes showing up as strong peaks in this RF spectrum. The displacement sensitivity of the sensor is then estimated based on the amplitudes and widths of these peaks above the photodetector-limited background and the expected thermal noise amplitude for motion based on the cantilever stiffness and mass [11, 12].

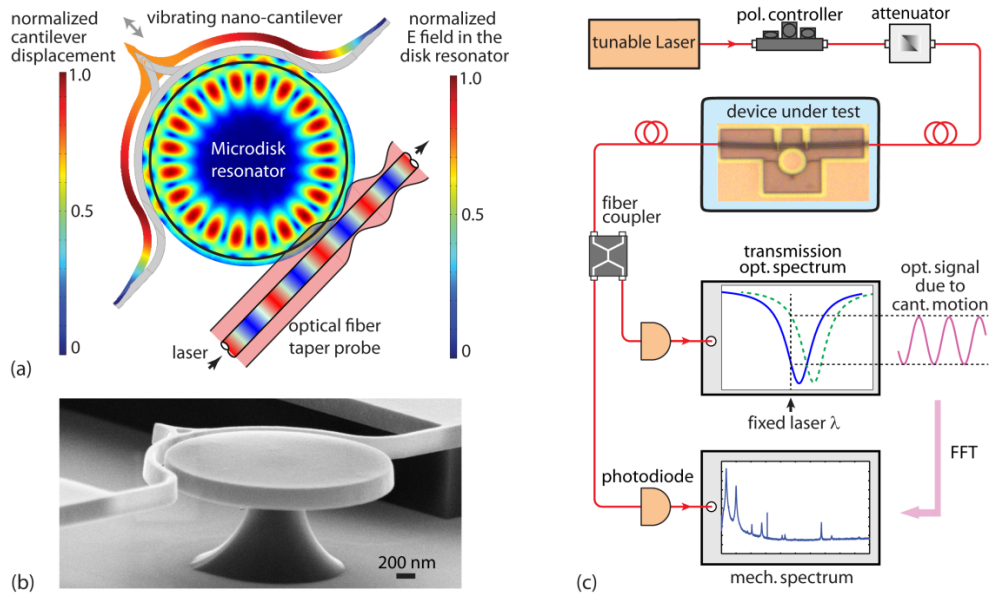

Fig. 1. (a) Working principle of the disk-cantilever device. The grey parts are the device at equilibrium. The colored cantilever shows the FEM-calculated deformed shape (with an exaggerated amplitude) of the first order, in-plane, even-symmetry mechanical mode, for a system with disk diameter of 2.5 μm, cantilever width of 125 nm, and cantilever thickness of 260 nm. The color map in the microdisk resonator represents the absolute value of the FEM-calculated electric field amplitude of the $TE_{1,10}$ optical mode. The mechanical motion of the cantilever is transduced by its influence on the microdisk optical mode, which is in- and out-coupled via the optical fiber taper waveguide probe. The left scale bar is for the cantilever displacement, the right one for the electric field amplitude in the microdisk, and the electric field in the fiber probe is not in scale. (b) Scanning electron micrograph of the fabricated device with nominal parameters the same as those in (a). (c) Schematic of the device characterization system.

Although there are many mechanical modes of the suspended cantilever, we focus on the first-order, in-plane, even-symmetry mode (see Fig. 1(a)), which is most strongly coupled to the microdisk optical modes and is of most relevance to our intended AFM application [3]. The ability of our system to optically transduce a given motional amplitude for this mechanical mode is a function of several parameters. The first is the optomechanical coupling parameter $g_{OM} = d\omega_c/dx$, given by the change in the microdisk mode's optical frequency $\omega_c$ per unit change in the disk-cantilever separation $x$. The optomechanical coupling determines how large a frequency shift is induced in the cavity's optical mode when the cantilever vibrates. The second parameter of importance is the microdisk's optical quality factor $Q$. The optical $Q$ determines the conversion between the frequency modulation of the microdisk optical mode created by the cantilever motion and the intensity modulation produced when the input laser is tuned to the shoulder of the microdisk optical mode. The final important parameters are the out-coupled power from the sensor and the noise equivalent power of the photodetector used to convert the optical signal to an electrical signal. The out-coupled power depends on the level of waveguide-microdisk coupling, all coupling losses within the system, and the input power, the last of which is limited in order to prevent processes like two-photon absorption and free carrier absorption (which are enhanced in microcavity geometries). The instrumental noise in the system is limited by the detector dark current, while the fundamental noise is imposed by the optical shot noise. Ideally, the out-coupled power is high enough that detection is shot-noise-limited.

Our sensor geometry has been chosen to optimize parameters such as $g_{OM}$ and $Q$. The semicircular cantilever shape increases the interaction length between the microdisk optical field and the cantilever's mechanical mode with respect to what can be achieved with a straight cantilever [10, 13, 14] (see Section 4.2 for more details). The semicircular cantilever shape largely preserves the low optical loss possible in Si microdisks [15], so that $Q \geq 10^4$ can be readily achieved. Finally, we note that while other cavity optomechanical systems [16, 17] support both higher $g_{OM}$ and $Q$ values, in most cases the mechanical structure involved has not been tailored to be a force/displacement transducer as needed in AFM applications. We have also observed both damping and regenerative mechanical oscillation in vacuum actuated by the gradient force, similar with that in Ref [16]. Furthermore, we anticipate that our design can be compatible with future integration with electrostatic actuation [18, 19] to drive the cantilever motion.

The goal in this work is to establish the cantilever stiffness range that can be effectively transduced using our device architecture, which, as described previously, has a large bearing on the range of AFM applications appropriate for this approach. Large changes in cantilever stiffness are achieved by adjusting the cantilever length, which in turn requires a change in the microdisk diameter in order to maintain strong optomechanical coupling. For a given cantilever length and disk diameter, further stiffness tuning is obtained by adjusting the cantilever width.

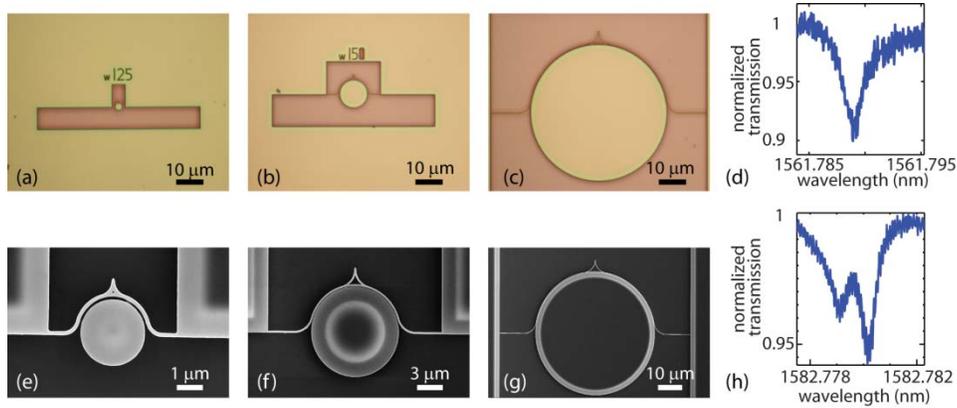

Fig. 2. Optical microscope images (a)-(c) and SEM images (e)-(g) of three fabricated devices. The disk diameter, $D$, and cantilever width, $w$, in the devices are: (a), (e) $D = 2.5$ μm, $w = 132$ nm ±6 nm; (b), (f) $D = 10$ μm, $w = 172$ nm ±5 nm; and (c), (g) $D = 50$ μm, $w = 155$ nm ±7 nm. Typical zoomed-in optical spectra of high-Q optical modes obtained from devices with (d) a 10 μm disk and (h) a 50 μm disk. The spectral widths and corresponding optical $Q$s are (d) ≈ 2.5 pm and ≈ 6.3×10$^5$, and (h) ≈ 0.9 pm and ≈ 1.8×10$^6$, respectively. All errors are one standard deviation unless stated otherwise.

## 3. Device fabrication and characterization setup

The devices were fabricated on silicon-on-insulator (SOI) wafers with a 260 nm top silicon device layer, 1 μm buried $SiO_2$ layer, and 625 μm thick Si handle wafer. A 360 nm thick positive electron beam (E-beam) resist was coated on the SOI wafer, followed by electron beam lithography. After development in hexyl acetate, the silicon device layer was etched through by an $SF_6/C_4F_8$ inductively-coupled plasma reactive ion etch with the patterned resist serving as the mask. After removing the resist in $O_2$ plasma, the $SiO_2$ layer was undercut by either 49 % HF or 6:1 buffered oxide etch depending on the disk diameter, followed by a liquid $CO_2$ critical point drying process, to release the silicon cantilever and the periphery of the disk without having the cantilever stick. Optical and scanning electron microscope (SEM) images of typical devices are shown in Fig. 2.

The characterization system employed an optical fiber taper waveguide [20] to evanescently couple light into and out of the microdisk resonator, as shown in Fig. 1(a) and (c). The fiber taper waveguide was fabricated by thinning the optical fiber down to ≈ 1 μm in diameter by heating and stretching. A local indentation ("dimple") with ≈ 10 μm radius of curvature is formed within the thinnest region of the fiber [21], allowing for selective probing of devices within two-dimensional arrays. A swept-wavelength laser with a wavelength range of 1520 nm to 1630 nm was used as the light source, and was sent into a polarization controller before going into the fiber taper waveguide, allowing for polarization adjustment to maximize the coupling depth of the desired optical mode before recording data. The device and the dimpled fiber probe were set up in a nitrogen environment (the blue box in Fig. 1(c)) to avoid device degradation (excess transmission loss of the fiber taper and stiction of the softest cantilevers due to moisture absorption). The output optical signal was split into two branches, with 10 % of the power sent to an InGaAs photodiode for the optical spectrum acquisition and 90 % to a 125 MHz bandwidth InGaAs photodiode for mechanical spectrum measurement. The typical optical powers at the input of the fiber taper waveguide were ≈ 15 μW and ≈ 100 μW for the measurements of optical and mechanical spectra, respectively.

## 4. Results and discussion

Devices with three different disk diameters (2.5 μm, 10 μm, and 50 μm; corresponding cantilever total lengths of ≈ 6 μm, ≈ 25 μm, and ≈ 105 μm; corresponding separation distances of the two cantilever clamps of 5 μm, 20 μm, and 80 μm, respectively) and several nominal cantilever widths (varying from 100 nm to 250 nm) per disk diameter were fabricated in order to achieve a cantilever stiffness range encompassing the ≈ 0.01 N/m to ≈ 100 N/m range used in most conventional AFM applications. The highest optical $Q$s are generally on the order of $10^5$ for devices with 10 μm disks and $10^6$ for those with 50 μm disks, as shown in Fig. 2(d) and (h). However, in many cases the highest $Q$ optical modes were not used for optomechanical transduction, for example, if the depth of coupling was too small (<10 %). Broader wavelength range optical spectra, the optical modes used in optomechanical transduction, and corresponding mechanical spectra from three typical devices are shown in Fig. 3.

The experimentally measured and calculated parameters of devices with different disk diameter, $D$, and cantilever width, $w$, are summarized in Table 1. The parameters were acquired in the following way. The mean values and standard deviations of $w$ were obtained from SEM images of the cantilevers, with six measurements along the cantilever for each $w$. The FEM simulation was used to calculate theoretical mechanical resonance frequency, $f_{mech}$, and the effective mass, $m_{eff}$, based on the measured $w$. $Q_{opt}$ is obtained from the experimental optical spectra and is the optical $Q$ of the mode used in transducing the mechanical motion. The experimental $f_{mech}$ and the mechanical $Q$, $Q_{mech}$, were obtained by fitting Lorentzian functions to the experimentally acquired mechanical spectra around the resonance peaks, since the displacement spectral density $S_{xx}(\omega)$ induced by thermal noise can be described by [11, 12]

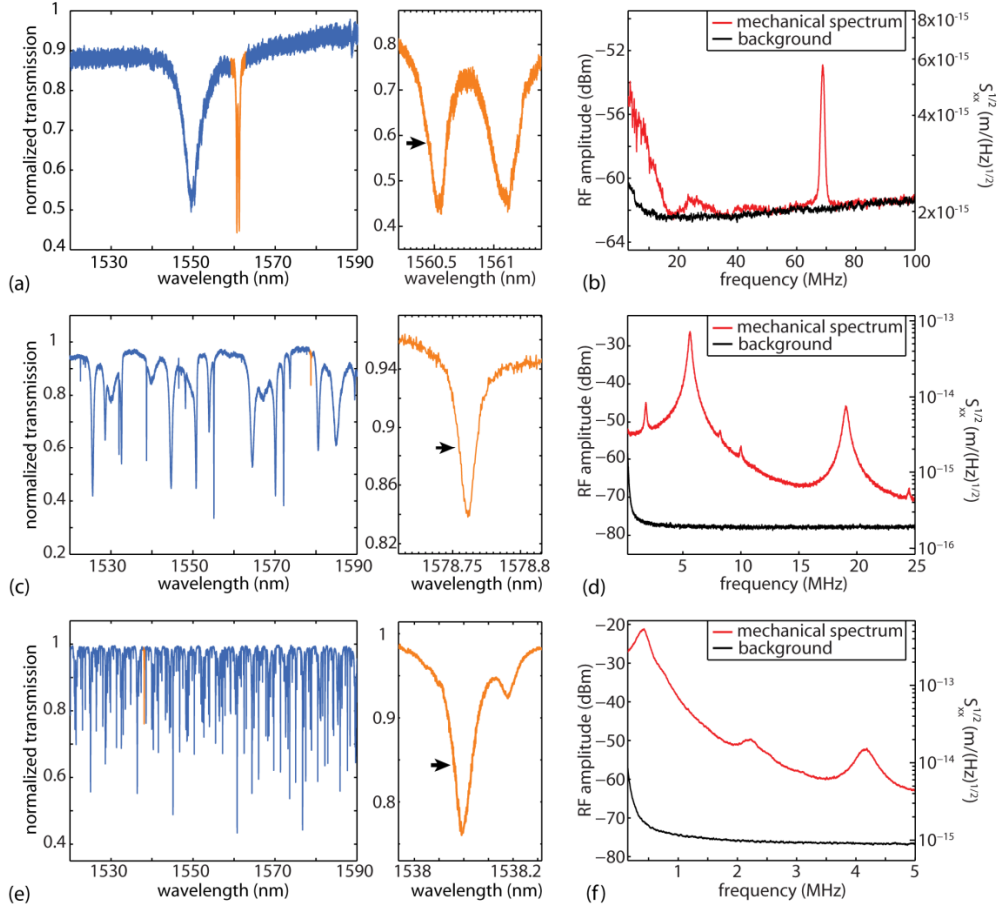

Fig. 3. Optical spectra (a), (c), and (e) and corresponding mechanical spectra (b), (d), and (f) of three typical disk-cantilever devices. The left figures in (a), (c), and (e) show the full-range optical spectra, with the optical modes used in transducing the corresponding mechanical spectra shown in the zoomed-in right figures (orange traces). The black arrows in the zoomed-in figures show the wavelengths used for mechanical spectrum measurements. The geometrical parameters of these devices are: (a), (b) $D$ = 2.5 μm, $w$ = 132 nm ±6 nm; (c), (d) $D$ = 10 μm, $w$ = 172 nm ±5 nm; and (e), (f) $D$ = 50 μm, $w$ = 155 nm ±7 nm.

$$S_{xx}(\omega) = \frac{4k_B T}{m_{eff}} \cdot \frac{\frac{\omega_m}{Q_{mech}}}{\left(\omega^2 - \omega_m^2\right)^2 + \frac{\omega^2 \omega_m^2}{Q_{mech}^2}}, \qquad (1)$$

where $k_B$ is the Boltzmann constant, $T$ the temperature (300 K), and $\omega_m = 2\pi f_{mech}$ the angular frequency. Using the experimentally obtained $f_{mech}$, we calculated the effective cantilever spring constant, $k_{eff} = m_{eff} \omega_m^2$, and the effective mean-square thermal displacement of the cantilever, $x_{rms} = \sqrt{k_B T / k_{eff}}$ [11, 12]. According to Eq. (1), the peak RF amplitude of the experimentally obtained mechanical spectra can be related with $x_{rms}$, allowing the calibration of $S_{xx}(\omega)$, which is shown in the right vertical axes in Fig. 3(b), (d), and (f).

The uncertainty in the width is the one standard deviation value from the SEM width measurements, and is listed in Table 1. The uncertainty in $m_{eff}$ due to that of the width is 10 %.

The uncertainty in the calculated $f_{mech}$ is 10 % and arises from the uncertainty in the moment of inertia due to the statistical error in width and the systematic error of the cross-section shape (see Section 4.5 for more detail), and the uncertainty in $m_{eff}$. The uncertainty in experimentally measured $f_{mech}$, $Q_{mech}$, and optical $Q$, given by the 95 % confidence intervals of the curve fitting are less than 1 % , 5 %, and 5 %, respectively. The uncertainties in $k_{eff}$ and $x_{rms}$ propagated from those in measured $f_{mech}$ and calculated $m_{eff}$ are 10 % and 5 %. The uncertainty of the displacement sensitivity arises from the combination of the systematic variation and the measurement uncertainty of the detector voltage noise power density and the uncertainty of the voltage to mechanical displacement calibration.

**Table 1.** Experimentally measured and calculated properties of the disk-cantilever devices. The typical photodetector-limited displacement sensitivity numbers are taken for representative devices within each disk diameter range ($D$ = 2.5 μm, $w$ = 205 nm, $D$ = 10 μm, $w$ = 172 nm, and $D$ = 50 μm, $w$ = 155 nm). The percentages listed for $f_{mech}$, $m_{eff}$, $k_{eff}$, $x_{rms}$, $Q_{mech}$, and the typical displacement sensitivity are the uncertainties in these quantities, and are described in the main text.

| $D$ (μm) | $w$ (nm) | $Q_{opt}$ ± 5 % | Exp. $f_{mech}$ (MHz) ± 1 % | Cal. $f_{mech}$ (MHz) ± 10 % | $m_{eff}$ (pg) ± 10 % | $k_{eff}$ (N/m) ± 10 % | $x_{rms}$ (pm) ± 5 % | $Q_{mech}$ ± 5 % | Typical disp sens. (fm/(Hz)$^{1/2}$) ± 15 % |
|---|---|---|---|---|---|---|---|---|---|
| 2.5 | 106±8 | 1.6×10$^4$ | 57.65 | 61 | 0.28 | 36 | 11 | 28 | |
| | 132±6 | 7.2×10$^3$ | 68.78 | 75 | 0.34 | 64 | 8.0 | 66 | |
| | 158±7 | 3.1 ×10$^3$ | 83.70 | 90 | 0.40 | 110 | 6.2 | 55 | 2.0 |
| | 205±7 | 5.2 ×10$^3$ | 100.0 | 120 | 0.51 | 200 | 4.5 | 69 | |
| | 238±12 | 1.2 ×10$^4$ | 111.4 | 140 | 0.60 | 290 | 3.8 | 80 | |
| 10 | 124±3 | 7.7×10$^4$ | 3.97 | 4.3 | 1.4 | 0.99 | 65 | 10 | |
| | 149±3 | 3.7×10$^4$ | 4.77 | 5.1 | 1.7 | 1.5 | 53 | 13 | |
| | 172±5 | 9.9×10$^4$ | 5.62 | 5.9 | 1.9 | 2.4 | 41 | 22 | 0.2 |
| | 224±3 | 1.4×10$^4$ | 7.30 | 7.7 | 2.5 | 5.3 | 28 | 18 | |
| | 256±4 | 8.7×10$^3$ | 8.17 | 8.8 | 2.9 | 7.6 | 23 | 21 | |
| | 271±5 | 1.5×10$^5$ | 8.87 | 9.3 | 3.0 | 9.4 | 21 | 27 | |
| 50 | 107±5 | 1.6×10$^4$ | 0.265 | 0.27 | 3.8 | 0.011 | 630 | 1.1 | |
| | 128±5 | 4.2×10$^4$ | 0.375 | 0.33 | 4.5 | 0.025 | 410 | 1.9 | |
| | 155±7 | 3.4×10$^4$ | 0.433 | 0.40 | 5.5 | 0.041 | 320 | 1.7 | 1.0 |
| | 210±5 | 3.4×10$^4$ | 0.538 | 0.54 | 7.4 | 0.085 | 220 | 2.5 | |
| | 233±10 | 6.2×10$^4$ | 0.615 | 0.60 | 8.2 | 0.12 | 180 | 4.3 | |

*4.1 Parameter range and comparison with conventional AFM*

According to the data shown in Fig. 4 and Table 1, we demonstrated that the motion of cantilevers with a stiffness ranging from ≈ 0.01 N/m to ≈ 292 N/m was effectively transduced through the optomechanical interaction with the adjacent microdisk optical cavity. A photodetector-limited displacement sensitivity as low as 0.2 fm/(Hz)$^{1/2}$ was achieved for a $D$ = 10 μm, $w$ = 172 nm device, which is slightly improved from the best results presented in Ref. [7]. More importantly, the displacement sensitivity remains in the fm/(Hz)$^{1/2}$ range across the full range of cantilever stiffness.

Part of the motivation in using nanoscale cantilevers for AFM is that, along with a wide accessible range of stiffness and fm/(Hz)$^{1/2}$ displacement sensitivity, mechanical resonance frequencies can be higher than in conventional AFM cantilevers. We observe that this is indeed the case in our devices - the fundamental mechanical frequencies (265 kHz to 111.4 MHz) are significantly higher than those of the conventional AFM probes (10 kHz to 1.1

MHz [8]). This high frequency range, enabled by the small mass of the nanoscale cantilevers, may increase the imaging acquisition rate, decrease thermal drifts [1], reduce ambient vibration and acoustic noise [8], and increase force sensitivity [22].

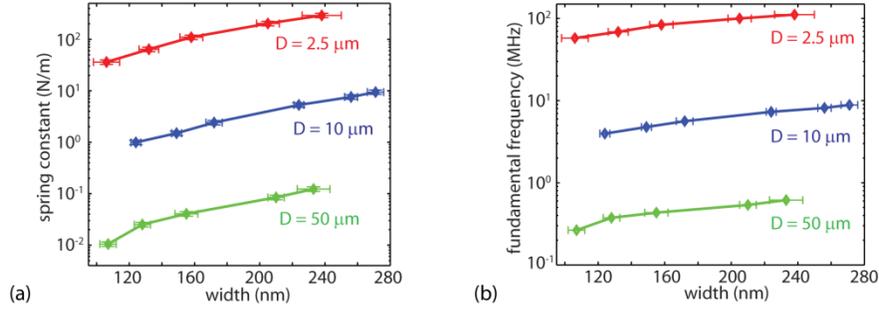

Fig. 4. (a) Spring constants and (b) experimentally measured fundamental mechanical frequencies of the fabricated devices. A spring constant range of over 4 orders of magnitude was achieved. The uncertainties in the fundamental mechanical frequency are not shown in (b) because they are much smaller than the data points.

The experimentally obtained and calculated $f_{mech}$ agree well, with a discrepancy between the two that is less than 10 % for all devices except for those with 2.5-µm disks and widths larger than 200 nm. One potential explanation for this is related to the clamping conditions assumed in the FEM simulations, where the cantilever is taken to be rigidly fixed at its two ends (See Fig. 1(a)). However, in the experimental case, since the cantilever is connected to the corner of an undercut silicon layer (See Fig. 1(b)), the ends of the cantilever may be able to move, rendering a lower stiffness (and mechanical frequency) than the calculated one. This is consistent with the calculated $f_{mech}$ being larger than the experimental value for all the devices with 2.5 µm or 10 µm disks, and with the discrepancy increasing with stiffness.

The mechanical $Q$ ($Q_{mech}$) of the devices mainly depends on the cantilever size and is limited by the viscous drag damping in the environment (which is at ambient pressure). The larger the disk size, the longer the cantilever and the larger viscous drag, resulting in lower $Q_{mech}$. The best $Q_{mech}$ achieved is around 80, less than the typical $Q$ (a few hundred [2]) of a conventional AFM. However, $Q_{mech}$ may be increased if the system is placed in vacuum. Preliminary measurements of 10 µm diameter disks with 100 nm width cantilevers show an improvement in $Q_{mech}$ by a factor of a few hundred, up to values as high as $5 \times 10^3$. Finally, we note that our semicircular cantilevers have been fabricated to include a sharp tip at their center point, with the idea being that this will eventually be used to image surface topography in AFM measurements. For this to be feasible, the tip radius should be smaller than the lateral size of the sample surface features that are to be resolved. The typical tip radius of our disk-cantilever devices is ≈ 10 nm (SEM image not shown); in comparison, micromachined silicon and silicon nitride tips of the conventional AFM have radii from <10 nm to 50 nm [1]. Our tip radius thus lies within the size range of the conventional AFM tips, while additional processing methods (focused ion beam milling or masked chemical etching) may be considered in the future to produce sharper tips.

*4.2 Optomechanical coupling parameter $g_{OM}$*

We next consider FEM calculations of the optomechanical coupling parameter $g_{OM}$ with a goal of understanding how it varies as the disk diameter (and cantilever length) changes in our devices. In the simulations, $g_{OM}$ was found in the following way. First, the mechanical modes of the cantilever were simulated to determine the frequency and shape of the mode of interest (in-plane even-symmetry mode). The cantilever was then deformed with the mechanical mode shape, until the gap between the disk and the center point of the cantilever reached the specific gap value. This was to simulate the real motion of the cantilever that we measured in the experiment. The resonant optical modes in the disk with the deformed cantilever were simulated by solving the eigenvalue problem of the optical field. After obtaining the resonant frequencies of a specific optical mode at different gaps, we find $g_{OM}$ as the slope of the fitted frequency-gap curve. We focus on the results obtained with 1st radial order optical modes, as these modes have the highest radiation-limited optical $Q$ and are predicted to couple well to the cantilever mechanical mode of interest. We define TE (TM) modes as ones with electric (magnetic) field components predominantly in the disk plane.

The simulation results of devices with different disk diameters and curved cantilevers are shown in Fig. 5. Simulation results of a device with a 10 μm disk and a straight cantilever are also included to illustrate the influence of the curved cantilever design. The straight cantilever (see Fig. 5(a)) has the same total length and width as the curved cantilever, and the disk-cantilever separation is taken to be the value at the point of the closest approach, which is the center point of the cantilever. We only show a TE mode for the 2.5 μm device in Fig. 5(b), because modes of TM polarization were not observed in the experiment. All the optical modes we studied in the simulation have wavelengths close to the ones used in the experiment. According to Fig. 5(b), at a gap of 100 nm, $g_{OM}/2\pi \approx 260$ MHz/nm for 50 μm devices, 550 MHz/nm for 10 μm devices, and 4.7 GHz/nm for 2.5-μm devices (TE modes). Further improvements can be made by reducing the gap to 50 nm, in which case the values increase to 750 MHz/nm, 2.14 GHz/nm, and 15.7 GHz/nm, respectively. In comparison, $g_{OM}/2\pi \approx 50$ MHz/nm at 100 nm gap for the same TE mode of 10 μm devices with straight cantilevers (dashed curves in Fig. 5(b)), $\approx 11$ times less than that with curved cantilevers. The $g_{OM}$ values with curved cantilevers are also consistently larger than those predicted in literature with geometries using straight cantilevers [10, 13], by one to two orders of magnitude. This improvement indicates one benefit of the curved cantilever design in terms of stronger optomechanical coupling, and the resulting potential for higher sensitivity in detection of the cantilever motion. The value of $g_{OM}$ decreases quickly when the gap increases from 30 nm to 200 nm, suggesting that a small gap is always beneficial for a large $g_{OM}$, if fabrication difficulties are not considered.

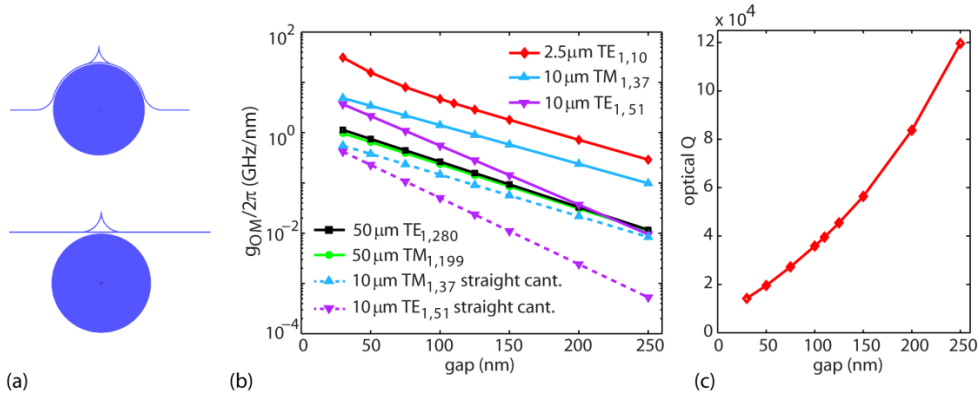

Fig. 5. (a) Sketches of 10 μm disks with (top) a curved cantilever and (bottom) a straight cantilever. (b) Simulated optomechanical coupling parameter $g_{OM}$ of TE/TM optical modes in the disk for devices with different disk sizes. Results for devices with curved cantilevers are shown as solid curves, while those for devices with straight cantilevers and 10 μm disks are presented as dashed curves for comparison. The cantilever width for all devices is set to 100 nm in the simulations. (c) Simulated optical $Q$ of the $TE_{1,10}$ mode for devices with 2.5 μm disk diameter, 100 nm cantilever width, and different cantilever-disk gaps.

Alongside $g_{OM}$, optical $Q$ is another key parameter which determines our system's displacement sensitivity. Whereas larger disks such as the 10 μm and 50 μm diameter devices support modes whose optical $Q$s are not severely degraded by the presence of the cantilever ($Q \geq 10^6$ in FEM simulations) [7], the 2.5 μm devices are predicted to have significantly limited $Q$s. In particular, while $Q \geq 10^6$ is predicted for a bare 2.5 μm disk, the introduction of the cantilever reduces this value, with $Q \approx 10^4$ to $10^5$ predicted for a gap of 100 nm (Fig. 5(c)). This is consistent with what was observed experimentally in Table 1. Thus, although $g_{OM}$ is significantly higher for these smaller diameter devices, the reduction in optical $Q$ serves as a mitigating factor that results in displacement sensitivities that remain similar to (or slightly poorer than) the smaller $g_{OM}$, but higher optical $Q$, 10 μm and 50 μm diameter devices.

*4.3 Fabrication considerations*

We have made progress in improving the quality and repeatability of our fabrication process relative to our previous work [7]. We now develop the E-beam resist at 8℃ instead of at room temperature, which makes the process less aggressive and helps create smoother edges of the developed features, resulting in a smoother sidewall of the disk resonator and cantilever. Another element to fabrication success is the consideration of the possible variation of the silicon dry etch rate over time and its reduced value within nanoscale gaps. To account for this, we determined the etch rate for each batch of devices by etching a test batch and cutting a cross-section through the disk-cantilever gap using focus ion beam milling. We also noticed the time lag between the development and the dry etch had an influence on the final feature size. After a 180 °C hotplate bake of the E-beam resist after spin-coating, we perform an additional 20 minute oven bake at 140 ℃ to remove any excess solvent and improve the resist film life and stability.

There are still some remaining issues and difficulties in the fabrication process. For example, the gap between the cantilever and disk was not always uniformly etched through. Another difficulty is that the cantilever is relatively frequently stuck in the devices with 50-μm disks, due to its low stiffness and long length, though we have found that a post-fabrication bake at 140 °C can help prevent eventual sticking problems. Overall, the average success rate in the fabricated devices was about 50 %, with a somewhat higher rate for the 2.5 μm diameter devices and a lower rate ($\approx$ 20 %) for the 50 μm devices.

*4.4 Factors limiting the stiffness range*

We believe we are approaching both the upper and the lower limits of the stiffness range that can be effectively transduced with the current disk-cantilever architecture and setup. On the soft side, the sticking issues discussed above make it increasingly difficult to obtain a freely suspended cantilever in air. We also fabricated 10 μm disks whose cantilevers had an extended length between the semicircular and clamped regions to decrease the stiffness, but the sticking issue remained. If the device is put into vacuum right after fabrication, smaller stiffness values might be possible to achieve. On the stiff side, the aforementioned limited optical $Q$s for 2.5 μm diameter devices, and the additional decrease in $Q$ with increasing cantilever width, suggest that further increase of the cantilever stiffness might require redesign of the device geometry.

*4.5 Influence of slanted sidewalls of the cantilever*

In fabricated devices, the cantilever sidewalls are sloped, resulting in a trapezoidal cross-section. The $w$ values listed in Table 1 were taken as the average of the top and the bottom widths as measured in the SEM. FEM simulations, on the other hand, were performed assuming that the cross-section of the cantilever is rectangular. To justify this simplification, we consider how the response of simple beams with rectangular and trapezoidal cross-sections varies. Since the deformation of the cantilever is mainly in-plane bending (See Fig. 1(a)), and the two cantilevers with different cross-sections have the same length, height, and mass, their behavior under the same excitation forces can be determined by the area moments of inertia, $I$. For a trapezoidal cross-section subjected to in-plane bending, [23]

$$I_{trap} = \frac{h}{36(w_1 + w_2)} \left[ w_1^4 + w_2^4 + 2w_1w_2\left(w_1^2 + w_2^2\right) \right.$$
$$\left. + \frac{w_2 - w_1}{2}\left(w_1^3 + 3w_1^2w_2 - 3w_1w_2^2 - w_2^3\right) + \left(\frac{w_2 - w_1}{2}\right)^2 \left(w_1^2 + 4w_1w_2 + w_2^2\right) \right], \quad (2)$$

where $w_1$ is the top width, $w_2$ the bottom width, $h$ the height of the cross-section. In the case of a rectangular cross-section with a width of $w = (w_1+w_2)/2$ and a height of $h$,

$$I_{rect} = \frac{w^3 h}{12}. \quad (3)$$

Because the difference between $w_1$ and $w_2$ is determined by the etch process, and does not depend on the average $w$, the difference between these two area moments of inertia is larger for smaller $w$. Two typical sets of data are: $w_1 = 70$ nm and $w_2 = 130$ nm with a smallest average $w$ of 100 nm, and $w_1 = 240$ nm and $w_2 = 297$ nm with a largest average $w$ of 269 nm. According to Eqs. (2) and (3), the difference between $I_{trap}$ and $I_{rect}$ are 8.3 % and 1.1 %, respectively, with $I_{trap}$ always larger than $I_{rect}$. This relatively small difference implies that the simulation results are reasonably reliable under the assumption of rectangular cross-sections.

*4.6 Outlook*

One might generically expect improved device performance (at least in terms of displacement sensitivity) if $g_{OM}$ and the optical $Q$ are increased. The most straightforward route to increased $g_{OM}$ without influencing the cantilever stiffness is through a reduction in the disk-cantilever gap (See Fig. 5(b)). While we have had some success in fabricating smaller gaps (Ref. [7]), this is not yet a repeatable process, and is made challenging due to the relatively large aspect ratio needed with the current E-beam resist process (>7:1 for a 50 nm gap). In addition, smaller gaps lead to lower optical $Q$ values at wider cantilever widths. Increasing the optical $Q$ is beneficial, provided that the frequency shift induced by the cantilever motion, $x$, is within the cavity mode line shape (i.e., the linear range of the optical readout approximately given by $g_{OM}x < \omega_c/Q$).

Achieving shot-noise-limited detection of the optically-transduced signal can be challenging within these silicon-based devices, as the amount of input power that can be injected into the microdisk cavity is limited by processes such as two-photon absorption and free-carrier absorption, which are enhanced in small mode volume, high quality factor cavities [24, 25]. As a result, one might consider optically amplifying the output signal from the system, or rather than direct photodetection, beating the output signal against a strong local oscillator in a homodyne measurement. In addition to allowing shot-noise-limited detection, such an approach can track the output amplitude and phase, the latter of which can be used as the dispersive signal needed to lock the input laser to the cavity.

For conventional AFM working in the dynamic mode, the tip oscillation amplitudes used in routine imaging range from a couple of nanometers to tens of nanometers. Although small amplitudes in the range of sub-nanometer help increase the sensitivity to short-range forces, the instability of the system discussed in Section 1 gets severe [2]. The thermally-driven displacement in our devices is not sufficiently large to be used in functional AFM applications. We can induce sufficiently large coherent cantilever oscillation in vacuum through radiation pressure forces, or alternatively using electrostatic or external piezoelectric actuation.

## 5. Conclusion

We experimentally demonstrated a cavity optomechanical sensor optimized for AFM applications, in which the motion of an integrated cantilever probe is optically transduced through near-field coupling to a microdisk optical cavity. Our main focus here was to demonstrate that this platform is compatible with a broad range of cantilever stiffness, and we demonstrate a stiffness range from $\approx$ 0.01 N/m to $\approx$ 290 N/m while maintaining fm/(Hz)$^{1/2}$ displacement sensitivity (or better). The resonance frequencies ($\approx$ 200 kHz to $\approx$ 110 MHz) of the cantilevers in our devices are much higher than those of conventional AFM, which may increase the imaging acquisition rate, reduce noise, and improve force sensitivity. This compact, monolithic sensor geometry and the large stiffness range that it can access suggest that these devices may have potential in a variety of AFM applications. With a low stiffness ($\approx$ 0.01 N/m), the device may be used as an on-chip version of AFM for label-free detection and in-situ characterization of microbiological samples [26]. With a high stiffness ($\approx$ 290 N/m), the small amplitude and increased sensitivity to short-ranged interaction forces may be suitable for high-resolution AFM imaging. Future work will focus on integration of electrostatic actuation of nanocantilever motion in this system, incorporation of these cavity optomechanical sensors in an AFM imaging setup, and in better understanding how techniques like radiation pressure cooling and excitation may be exploited in AFM measurements.

## Acknowledgements

The authors thank Richard Kasica of the CNST NanoFab for assistance with the development of E-beam lithography process, and Marcelo Davanço for help with FEM simulations. Yuxiang Liu acknowledges support under the National Institute for Standards and Technology American Recovery and Reinvestment Act Measurement Science and Engineering Fellowship Program Award 70NANB10H026 through the University of Maryland. Houxun Miao acknowledges support under the Cooperative Research Agreement between the University of Maryland and the National Institute of Standards and Technology Center for Nanoscale Science and Technology, Award 70NANB10H193, through the University of Maryland.